\documentclass[twocolumn,aps,prl,preprintnumbers]{revtex4-2}
\usepackage{bbm}
\usepackage[latin9]{inputenc}
\usepackage{float}
\usepackage{amsmath}
\usepackage{amssymb}
\usepackage{graphicx}
\usepackage{mathrsfs}
\usepackage{amsfonts}
\usepackage{amsthm}
\usepackage{color}
\usepackage{txfonts}
\usepackage[colorlinks=true,citecolor=blue,linkcolor=blue,urlcolor=blue,anchorcolor=blue]{hyperref}%
\usepackage{pifont}
\usepackage{multirow}
\usepackage{booktabs}
\hypersetup{colorlinks=true,citecolor=blue,linkcolor=blue,urlcolor=blue}
\providecommand{\U}[1]{\protect\rule{.1in}{.1in}}
\makeatletter
\@ifundefined{textcolor}{}
{
\definecolor{BLACK}{gray}{0}
\definecolor{WHITE}{gray}{1}
\definecolor{RED}{rgb}{1,0,0}
\definecolor{GREEN}{rgb}{0,1,0}
\definecolor{BLUE}{rgb}{0,0,1}
\definecolor{CYAN}{cmyk}{1,0,0,0}
\definecolor{MAGENTA}{cmyk}{0,1,0,0}
\definecolor{YELLOW}{cmyk}{0,0,1,0}
}

\makeatother

\begin{document}
\title{Instability of two-dimensional nonrelativistic altermagnets}
\author{Zhejunyu Jin$^{1,2}$ and Peng Yan$^{1}$}
\email[Contact author: ]{yan@uestc.edu.cn}
\affiliation{$^1$School of Physics and State Key Laboratory of Electronic Thin Films and Integrated Devices, University of
Electronic Science and Technology of China, Chengdu 610054, China\\
$^2$Institute of Solid State Theory, University of M\"{u}nster, D-48149 M\"{u}nster, Germany}

\begin{abstract}
Altermagnets (AMs) are collinear compensated magnets that exhibit momentum-dependent spin splitting without requiring a net magnetization. Since local magnetic moments and exchange-driven order do not rely on relativistic effects, altermagnetism can be naturally formulated in a nonrelativistic spin-group framework. This raises a basic question: can two-dimensional altermagnetic order be stabilized by purely exchange interactions in the absence of spin-orbit-induced anisotropy? We address this question by applying Bogoliubov's inequality to a minimal $d$-wave altermagnetic spin model. We show that the anisotropic exchange pattern responsible for altermagnetism still contributes only a quadratic long-wavelength term to the Bogoliubov denominator. Consequently, short-ranged exchange interactions alone cannot stabilize long-range altermagnetic order in two dimensions when continuous spin-rotation symmetry is preserved. In contrast, the corresponding three-dimensional system has a finite small-momentum contribution and is not ruled out by the Mermin-Wagner argument.
\end{abstract}

\maketitle

\textit{Introduction---}Altermagnets (AMs) represent an emerging class of magnet that goes beyond the conventional distinction between ferromagnets and antiferromagnets \cite{Smejkal3}. Like antiferromagnets, AMs possess compensated magnetic order and zero net magnetization. Unlike conventional antiferromagnets, however, they break the combined inversion and time-reversal symmetry and exhibit momentum-dependent spin splitting even in the absence of spin-orbit coupling \cite{Jungwirth2025,Smejkal1,Hayami2020,McClarty2024,Gomonay2024,Ma2021,Duan2025,Osumi2024,Jiang2025,Smejkal2,Cui2023,Jin2026,Yang2026,Neumann2026}. The nonrelativistic description of AMs is naturally formulated in terms of spin groups \cite{Litvin1974,Litvin1977}, which classify altermagnets into distinct symmetry classes, such as $d$-, $g$-, and $i$-wave types \cite{Smejkal1}. In this framework, opposite-spin sublattices are related by crystal rotations or mirror operations combined with spin rotations. These symmetry considerations establish altermagnetism as a nonrelativistic magnetic phase distinct from conventional antiferromagnetism, but they do not by themselves determine whether the corresponding long-range order is stable in low dimensions.

Originally formulated for isotropic Heisenberg ferromagnets and antiferromagnets, the Mermin-Wagner theorem states that a continuous spin symmetry cannot be spontaneously broken at finite temperature in one- or two-dimensional systems with sufficiently short-ranged exchange interactions \cite{Mermin1966}. Equivalently, conventional ferromagnetic or antiferromagnetic long-range order is forbidden in strictly one- and two-dimensional isotropic spin systems. This observation naturally raises the central question addressed here: does the same Mermin-Wagner long-wavelength constraint apply to two-dimensional altermagnets? In particular, can an altermagnet whose magnetic order and spin splitting are generated purely by nonrelativistic anisotropic exchange interactions sustain long-range magnetic order at finite temperature in the absence of spin-orbit-induced anisotropy?
 
\begin{figure}
    \centering
    \includegraphics[width=\columnwidth]{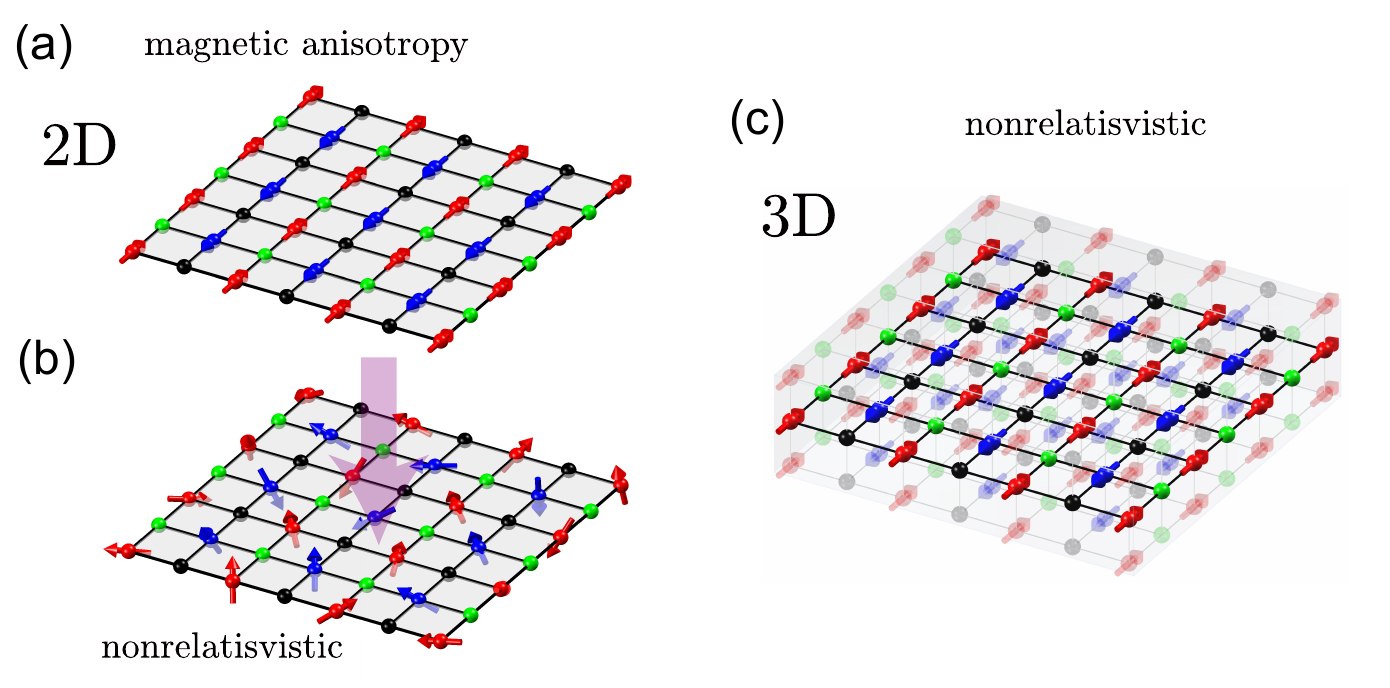}
    \caption{Schematic summary of the Mermin-Wagner argument for altermagnets. A two-dimensional altermagnet can be stabilized by magnetic anisotropy (a), whereas long-range magnetic order is destroyed in the purely nonrelativistic exchange limit with continuous spin-rotation symmetry (b). In three dimensions, the Mermin-Wagner argument does not forbid exchange-driven altermagnetic order (c). Red and blue spheres denote magnetic atoms, while green and black spheres denote inequivalent nonmagnetic atoms.}
    \label{fig1}
\end{figure}

In this work, we examine this issue by adapting the original Mermin-Wagner argument to altermagnets. We show that the anisotropic exchange pattern responsible for $d$-wave altermagnetism still produces only a $k^2$ contribution to the Bogoliubov denominator. Consequently, purely exchange-driven two-dimensional AMs are unstable at finite temperature when continuous spin-rotation symmetry is preserved. In three dimensions, the corresponding small-momentum integral remains finite, so the Mermin-Wagner argument no longer excludes long-range altermagnetic order, as summarized in Fig.~\ref{fig1}. Our result clarifies that altermagnetic spin splitting by itself does not provide the symmetry-breaking gap needed to evade the Mermin-Wagner theorem; magnetic anisotropy or another spin-symmetry-breaking mechanism is required to stabilize two-dimensional altermagnetic order.

\textit{Results---}We start from Bogoliubov's inequality \cite{Wagner1966}
\begin{equation}\label{Eq1}
\begin{aligned}
\frac{1}{2}\langle\{D,D^{\dagger}\}\rangle \langle[[C,H],C^{\dagger}]\rangle\geq k_B T|\langle[C,D]\rangle|^2,
\end{aligned}
\end{equation}
where $H$ is the Hamiltonian and $\langle X\rangle={\rm Tr} (Xe^{-\beta H})/{\rm Tr} (e^{-\beta H})$ with $\beta=1/k_B T$. Here, $T$ is the absolute temperature and $k_B$ is the Boltzmann constant.

We apply this inequality to a two-dimensional $d$-wave altermagnet described by the spin Hamiltonian
\begin{equation}\label{Eq2}
\begin{aligned}
H=H_{A}+H_{B}+H_{AB}+H_h+H_a,
\end{aligned}
\end{equation}
with
\begin{equation}\label{Eq3}
\begin{aligned}
H_A&=-\sum_{{\bf R}_A,{\bf R}'_A}J_A({\bf R}_A-{\bf R}'_A){\bf S}({\bf R}_A)\cdot {\bf S}({\bf R}'_A),\\
H_B&=-\sum_{{\bf R}_B,{\bf R}'_B}J_B({\bf R}_B-{\bf R}'_B){\bf S}({\bf R}_B)\cdot {\bf S}({\bf R}'_B),\\
H_{AB}&=-\sum_{{\bf R}_A,{\bf R}_B}J_{AB}({\bf R}_A-{\bf R}_B){\bf S}({\bf R}_A)\cdot {\bf S}({\bf R}_B),\\
H_h&=-h\sum_{\bf R} S_z ({\bf R}) e^{-i{\bf K}\cdot{\bf R}},\\
H_a&=-K_0\sum_{\bf R}S_z^2({\bf R}).
\end{aligned}
\end{equation}
We consider a Lieb-lattice realization of the $d$-wave altermagnet \cite{Durrnagel2025,Kaushal2025,Cui2023}. The two magnetic sublattices are denoted by $A$ and $B$, with ${\bf R}_A$ (${\bf R}_B$) labeling a site on the $A$ ($B$) sublattice, and ${\bf S}({\bf R}_{A,B})$ being the spin operator at that site. A coordinate ${\bf R}$ without a sublattice label denotes a generic magnetic site. The terms $H_A$, $H_B$, and $H_{AB}$ describe exchange interactions within the $A$ sublattice, within the $B$ sublattice, and between the two sublattices, respectively. The next-nearest-neighbor exchange is anisotropic and commutative on the two sublattices:
\begin{equation}\label{Eq3a}
\begin{aligned}
J_A(\pm a\hat{x})&=J_1,& J_A(\pm a\hat{y})&=J_2,\\
J_B(\pm a\hat{x})&=J_2,& J_B(\pm a\hat{y})&=J_1,
\end{aligned}
\end{equation}
while the nearest-neighbor exchange between the two sublattices is assumed to be isotropic, $J_{AB}=J_0$. Here, $a$ is the in-plane lattice constant, i.e., the distance between same-sublattice next-nearest neighbors along $\hat{x}$ or $\hat{y}$. The external field $h$ couples to the order parameter with ordering wave vector ${\bf K}$. We choose the phase convention such that $e^{-i{\bf K}\cdot{\bf R}}=1$ for sites belonging to the same spin sublattice and $e^{-i{\bf K}\cdot{\bf R}}=-1$ for sites belonging to opposite spin sublattices. The parameter $K_0$ denotes the single-ion anisotropy strength. We take $C=S_{+}({\bf k})$ and $D=S_{-}(-{\bf k}-{\bf K})$, with wavevector $\mathbf{k}$ and
\begin{equation}\label{Eq4}
\begin{aligned}
S_{\pm}({\bf k})=\frac{1}{\sqrt{2}}\sum_{\bf R}e^{-i{\bf k}\cdot{\bf R}}
\left[S_x({\bf R})\pm iS_y({\bf R})\right].
\end{aligned}
\end{equation}
For these operators,
\begin{equation}\label{Eq4a}
\begin{aligned}
\langle[C,D]\rangle
&=\sum_{\bf R}e^{i{\bf K}\cdot{\bf R}}\langle S_z({\bf R})\rangle\equiv Ns_z,\\
\frac{1}{2}\sum_{\bf k}\langle\{D,D^{\dagger}\}\rangle
&=\frac{N}{2}\sum_{\bf R}\langle S_x^2({\bf R})+S_y^2({\bf R})\rangle
\leq \frac{N^2}{2}S(S+1),
\end{aligned}
\end{equation}
where $s_z=N^{-1}\sum_{\bf R}e^{i{\bf K}\cdot{\bf R}}\langle S_z({\bf R})\rangle$ is the staggered magnetization density, $N$ is the number of spins, and $S$ is the spin quantum number, so that ${\bf S}^2=S(S+1)$. The second relation is the full-Brillouin-zone anticommutator bound and therefore also bounds any restricted momentum sum.
Using the commutation relations
\begin{equation}\label{Eq5}
\begin{aligned}
\relax [S_i({\bf R}),S_j({\bf R}')]&=i\epsilon_{ijk}S_k({\bf R})\delta_{{\bf R},{\bf R}'},\\
[S_+({\bf R}),S_-({\bf R}')]&=S_z({\bf R})\delta_{{\bf R},{\bf R}'}, \\
\end{aligned}
\end{equation}
where $\epsilon_{ijk}$ is the Levi-Civita tensor with $i,j,k=x,y,z$, one obtains
\begin{equation}\label{Eq6}
\begin{aligned}
\relax [C,H_A]
=&\frac{1}{\sqrt{2}}\sum_{{\bf R}_A,{\bf R}'_A}J_A({\bf R}_A-{\bf R}'_A)\bigg\{
iS_z({\bf R}_A)S_y({\bf R}'_A)(e^{-i{\bf k}\cdot{\bf R}'_A}-e^{-i{\bf k}\cdot{\bf R}_A})\\
&+iS_y({\bf R}_A)S_z({\bf R}'_A)(e^{-i{\bf k}\cdot{\bf R}_A}-e^{-i{\bf k}\cdot{\bf R}'_A})\\
&+S_x({\bf R}_A)S_z({\bf R}'_A)(e^{-i{\bf k}\cdot{\bf R}_A}-e^{-i{\bf k}\cdot{\bf R}'_A})\\
&+S_z({\bf R}_A)S_x({\bf R}'_A)(e^{-i{\bf k}\cdot{\bf R}'_A}-e^{-i{\bf k}\cdot{\bf R}_A})\bigg\}.
\end{aligned}
\end{equation}
After collecting the terms, the double commutator is
\begin{equation}\label{Eq7}
\begin{aligned}
\relax [[C,H_A],C^\dagger]
=&\sum_{{\bf R}_A,{\bf R}'_A}J_A({\bf R}_A-{\bf R}'_A)
(e^{-i{\bf k}\cdot{\bf R}_A}-e^{-i{\bf k}\cdot{\bf R}'_A})\\
&\times\bigg[
(e^{i{\bf k}\cdot{\bf R}_A}-e^{i{\bf k}\cdot{\bf R}'_A})S_z({\bf R}_A)S_z({\bf R}'_A)\\
&+e^{i{\bf k}\cdot{\bf R}_A}S_-({\bf R}_A)S_+({\bf R}'_A)\\
&-e^{i{\bf k}\cdot{\bf R}'_A}S_+({\bf R}_A)S_-({\bf R}'_A)\bigg].
\end{aligned}
\end{equation} 
Similar calculations give
\begin{equation}\label{Eq8}
\begin{aligned}
\relax [[C,H_B],C^\dagger]
=&\sum_{{\bf R}_B,{\bf R}'_B}J_B({\bf R}_B-{\bf R}'_B)
(e^{-i{\bf k}\cdot{\bf R}_B}-e^{-i{\bf k}\cdot{\bf R}'_B})\\
&\times\bigg[
(e^{i{\bf k}\cdot{\bf R}_B}-e^{i{\bf k}\cdot{\bf R}'_B})S_z({\bf R}_B)S_z({\bf R}'_B)\\
&+e^{i{\bf k}\cdot{\bf R}_B}S_-({\bf R}_B)S_+({\bf R}'_B)\\
&-e^{i{\bf k}\cdot{\bf R}'_B}S_+({\bf R}_B)S_-({\bf R}'_B)\bigg],
\end{aligned}
\end{equation}
and 
\begin{equation}\label{Eq9}
\begin{aligned}
\relax [[C,H_{AB}],C^\dagger]
=&\sum_{{\bf R}_A,{\bf R}_B}J_{AB}({\bf R}_A-{\bf R}_B)
(e^{-i{\bf k}\cdot{\bf R}_A}-e^{-i{\bf k}\cdot{\bf R}_B})\\
&\times\big[
(e^{i{\bf k}\cdot{\bf R}_A}-e^{i{\bf k}\cdot{\bf R}_B})S_z({\bf R}_A)S_z({\bf R}_B)\\
&+e^{i{\bf k}\cdot{\bf R}_A}S_-({\bf R}_A)S_+({\bf R}_B)\\
&-e^{i{\bf k}\cdot{\bf R}_B}S_+({\bf R}_A)S_-({\bf R}_B)\big].
\end{aligned}
\end{equation} 
To bound the exchange contribution to the double commutator in Eqs.~(\ref{Eq7})--(\ref{Eq9}), we have
\begin{equation}\label{Eq10}
\begin{aligned}
2\sum_{\bf R}
\left\langle S_-({\bf R}+{\bf d})S_+({\bf R})
+S_z({\bf R}+{\bf d})S_z({\bf R})\right\rangle\leq 2N S(S+1).
\end{aligned}
\end{equation}
The inequality follows from Schwarz's inequality and the local operator bound
$S_z^2+\frac{1}{2}(S_x^2+S_y^2)\leq S(S+1)$. For the Lieb lattice considered here, the intra-sublattice next-nearest-neighbor bonds lie along $\pm a\hat{x}$ and $\pm a\hat{y}$. The nearest-neighbor $A$-$B$ bonds are described by the displacement vectors ${\boldsymbol\delta}=(\pm a/2,\pm a/2)$, whose length is $a/\sqrt{2}$. Consequently, we obtain
\begin{equation}\label{Eq11}
\begin{aligned}
\langle[[C,H_A],C^\dagger]\rangle &\leq N S(S+1)a^2 (|J_1|k_x^2+|J_2|k_y^2),\\
\langle[[C,H_B],C^\dagger]\rangle &\leq N S(S+1)a^2 (|J_2|k_x^2+|J_1|k_y^2),\\
\langle[[C,H_{AB}],C^\dagger]\rangle &\leq N S(S+1)\frac{1}{2}a^2 |J_0|(k_x^2+k_y^2),
\end{aligned}
\end{equation}
where we have used $1-\cos(ka)\leq\frac{1}{2} (ka)^2$. Therefore, one obtains
\begin{equation}\label{Eq12}
\begin{aligned}
\langle[[C,H_A+H_B+H_{AB}],C^\dagger]\rangle
&\leq N J' k^2,
\end{aligned}
\end{equation}
with
\begin{equation}\label{Eq13}
\begin{aligned}
J'=S(S+1)a^2(|J_1|+|J_2|+\frac{1}{2}|J_0|).
\end{aligned}
\end{equation}
We next consider the contribution from the staggered field,
\begin{equation}\label{Eq14}
\begin{aligned}
\relax [C,H_h]&=[S_{+}({\bf k}),-h\sum_{\bf R}S_z({\bf R})e^{-i{\bf K}\cdot{\bf R}}]\\
&=hS_+({\bf k}+{\bf K}).
\end{aligned}
\end{equation}
We thus obtain
\begin{equation}\label{Eq15}
\begin{aligned}
\langle[[C,H_h],C^{\dagger}]\rangle
&=h\sum_{\bf R}\langle S_z({\bf R})e^{-i{\bf K}\cdot{\bf R}}\rangle
=hNs_z.
\end{aligned}
\end{equation}
Finally, we consider the contribution from the uniaxial single-ion anisotropy.
\begin{equation}\label{Eq16}
\begin{aligned}
\relax [S_x({\bf R}),S_z^2({\bf R})]
&=[S_x({\bf R}),S_z({\bf R})]S_z({\bf R})
+S_z({\bf R})[S_x({\bf R}),S_z({\bf R})]\\
&=-iS_y({\bf R})S_z({\bf R})-iS_z({\bf R})S_y({\bf R})\\
&=S_x({\bf R})-2iS_z({\bf R})S_y({\bf R}),
\end{aligned}
\end{equation}
and
\begin{equation}\label{Eq17}
\begin{aligned}
i[S_y({\bf R}),S_z^2({\bf R})]
&=i\left([S_y({\bf R}),S_z({\bf R})]S_z({\bf R})
+S_z({\bf R})[S_y({\bf R}),S_z({\bf R})]\right)\\
&=-S_x({\bf R})S_z({\bf R})-S_z({\bf R})S_x({\bf R})\\
&=iS_y({\bf R})-2S_z({\bf R})S_x({\bf R}).
\end{aligned}
\end{equation}
Combining Eqs.~(\ref{Eq16}) and (\ref{Eq17}) gives
\begin{equation}\label{Eq18}
\begin{aligned}
\relax [C,H_a]
&=-K_0\sum_{\bf R}e^{-i{\bf k}\cdot{\bf R}}
[S_+({\bf R}),S_z^2({\bf R})]\\
&=K_0\sum_{\bf R}e^{-i{\bf k}\cdot{\bf R}}
\left[2S_z({\bf R})-1\right]S_+({\bf R}),
\end{aligned}
\end{equation}
and
\begin{equation}\label{Eq19}
\begin{aligned}
\left[\left(2S_z({\bf R})-1\right)S_+({\bf R}),S_-({\bf R})\right]
&=2S_z^2({\bf R})-S_x^2({\bf R})-S_y^2({\bf R})\\
&=3S_z^2({\bf R})-S(S+1).
\end{aligned}
\end{equation}
Consequently, we have
\begin{equation}\label{Eq20}
\begin{aligned}
\langle [[C,H_a],C^\dagger]\rangle
&=K_0\sum_{\bf R}
\left\langle 3S_z^2({\bf R})-S(S+1)\right\rangle .
\end{aligned}
\end{equation}
For $K_0>0$, this term is bounded by
\begin{equation}\label{Eq21}
\begin{aligned}
\langle [[C,H_a],C^\dagger]\rangle
&\leq K_0N\left[3S^2-S(S+1)\right]
=K_0N(2S^2-S).
\end{aligned}
\end{equation}

\begin{figure}
    \centering
    \includegraphics[width=\columnwidth]{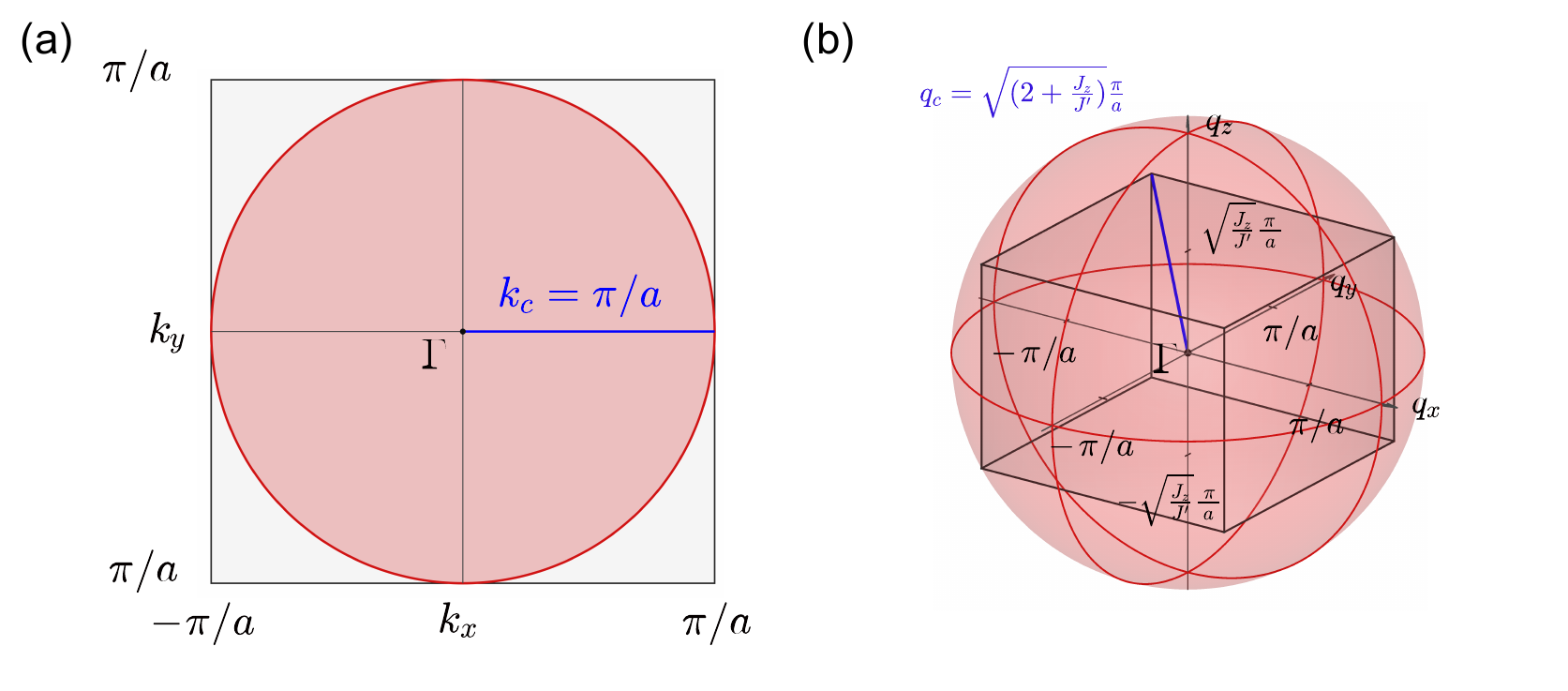}
    \caption{Schematic of the momentum cutoffs used in the long-wavelength analysis. The left panel shows the first Brillouin zone of the two-dimensional square Bravais lattice and the disk $|{\bf k}|<k_c$, which gives a lower bound for the two-dimensional integral. The right panel shows the image of the three-dimensional Brillouin-zone cutoff region in the rescaled momentum variables $q_x=k_x$, $q_y=k_y$, and $q_z=\sqrt{J_z/J'}\,k_z$, together with a spherical cutoff $|{\bf q}|<q_c$ enclosing this region. The plot illustrates the case $J_z<J'$ with $c=a$. The corner markers lie on the enclosing sphere, and the blue segment in panel (b) points to one such corner, whose distance from the origin is $q_c$.}
    \label{fig2}
\end{figure}

For $S=1/2$, $(2S^2-S)=0$, so the single-ion term gives no contribution to the double commutator. We first consider the two-dimensional $d$-wave altermagnet. For each momentum, Eq.~(\ref{Eq1}) and the bound $\langle[[C,H],C^\dagger]\rangle\leq N(J'k^2+B)$ give
\begin{equation}\label{Eq21a}
\begin{aligned}
\frac{1}{2}\langle\{D,D^\dagger\}\rangle
&\geq \frac{k_{\rm B}TNs_z^2}{J'k^2+B}.
\end{aligned}
\end{equation}
Summing Eq.~(\ref{Eq21a}) over the low-momentum disk and using the anticommutator bound in Eq.~(\ref{Eq4a}) gives
\begin{equation}\label{Eq21b}
\begin{aligned}
\frac{N^2}{2}S(S+1)
&\geq k_{\rm B}TNs_z^2
\sum_{|{\bf k}|\leq k_c}\frac{1}{J'k^2+B}.
\end{aligned}
\end{equation}
Using $\sum_{\bf k}\to (N/\rho)\int d^2k/(2\pi)^2$, the factors of $N$ cancel and we obtain
\begin{equation}\label{Eq22}
\begin{aligned}
S(S+1)\geq \frac{2k_{\rm B}Ts_z^2}{\rho}
\int_{|{\bf k}|\leq k_c}\frac{d^2 k}{(2\pi)^2}
\frac{1}{J'k^2+B}.
\end{aligned}
\end{equation}
Here, $k_c$ can be chosen as the distance from the origin to the nearest Bragg plane. For the square Bravais lattice, $k_c=\pi/a$, as shown in Fig.~\ref{fig2}. The disk $|{\bf k}|<k_c$ does not replace the entire Brillouin zone; rather, it is a low-momentum region contained inside it. Since the integrand is positive, the integral gives a lower bound to the full Brillouin-zone contribution. Therefore, if the disk contribution diverges, the full integral must also diverge. $B=|hs_z|+K_0(2S^2-S)$ is induced by the staggered field and magnetic anisotropy. It breaks the continuous spin-rotation symmetry and gaps the Goldstone mode. $\rho$ is the spin density, i.e., the number of spins per unit area in two dimensions and per unit volume in three dimensions \cite{Mermin1966}. The integral on the right-hand side of (\ref{Eq22}) is
\begin{equation}\label{Eq23}
\begin{aligned}
I_{2D}=\int_{|{\bf k}|\leq k_c}\frac{d^2 k}{(2\pi)^2}\frac{1}{J'k^2+B}&=\frac{1}{2\pi}\int_{0}^{k_c} dk\frac{k}{J'k^2+B}\\
&=\frac{1}{4\pi J'}\ln(\frac{J'k_c^2+B}{B}).
\end{aligned}
\end{equation}
In the absence of the staggered field and magnetic anisotropy, $B\to0$. The two-dimensional integral then diverges logarithmically, which is incompatible with a nonzero staggered magnetization at finite temperature.

We next consider the three-dimensional case, with an additional interlayer exchange interaction
\begin{equation}\label{Eq24}
\begin{aligned}
H_{\rm inter}=-\sum_{{\bf R}_z,{\bf R}'_z}J_3{\bf S}({\bf R}_z)\cdot {\bf S}({\bf R}'_z).
\end{aligned}
\end{equation}
Here, ${\bf R}_z$ and ${\bf R}'_z$ denote magnetic sites in neighboring layers connected by the interlayer exchange $J_3$, and the interlayer spacing is denoted by $c$. We then have
\begin{equation}\label{Eq25}
\langle[[C,H_{\rm inter}],C^\dagger]\rangle \leq N J_z k_z^2,
\end{equation}with $J_z\equiv S(S+1)c^2 |J_3|$.

The corresponding inequality contains the three-dimensional integral
\begin{equation}\label{Eq26}
\begin{aligned}
S(S+1)\geq \frac{2k_{\rm B}Ts_z^2}{\rho}
\int \frac{d^3 k}{(2\pi)^3}
\frac{1}{J'(k_x^2+k_y^2)+J_z k_z^2+B}.
\end{aligned}
\end{equation}
By defining
\begin{equation*}
\begin{aligned}
q_x=k_x,\qquad
q_y=k_y,\qquad
q_z=\sqrt{\frac{J_z}{J'}}\,k_z,
\end{aligned}
\end{equation*}
the denominator becomes $J'(q_x^2+q_y^2+q_z^2)+B$. Under this change of variables, the first Brillouin zone in ${\bf k}$ space is mapped to a rescaled region $\mathcal E_q$ in ${\bf q}$ space. To show that the three-dimensional integral is finite, we enclose $\mathcal E_q$ by a sphere of radius $q_c$:
\begin{equation}\label{Eq27}
\begin{aligned}
I_{3D}&=\int_{\rm BZ} \frac{d^3 k}{(2\pi)^3}\frac{1}{J'(k_x^2+k_y^2)+J_z k_z^2+B}\\
&=\frac{1}{\sqrt{J'J_z}}\int_{\mathcal E_q}\frac{d^3 q}{(2\pi)^3}\frac{1}{q^2+B/J'}\\
&\leq\frac{1}{2\pi^2\sqrt{J'J_z}}\int_0^{q_c}dq\frac{q^2}{q^2+B/J'}\\
&=\frac{1}{2\pi^2\sqrt{J'J_z}}\bigg[q_c-\sqrt{\frac{B}{J'}}\arctan\left(\frac{q_c}{\sqrt{B/J'}}\right)\bigg].
\end{aligned}
\end{equation}
Here, $q=|{\bf q}|=\sqrt{q_x^2+q_y^2+q_z^2}$, and $q_c$ is the radius of the spherical cutoff enclosing the rescaled momentum region, as shown in Fig.~\ref{fig2} (b). If the original Brillouin-zone boundaries are $|k_x|,|k_y|<\pi/a$ and $|k_z|<\pi/c$, one may choose
\begin{equation}\label{Eq28}
\begin{aligned}
q_c=\sqrt{2\left(\frac{\pi}{a}\right)^2+\frac{J_z}{J'}\left(\frac{\pi}{c}\right)^2}.
\end{aligned}
\end{equation}
For $c=a$, this gives $q_c=(\pi/a)\sqrt{2+J_z/J'}$. The precise value of $q_c$ is cutoff dependent and does not affect the small-momentum conclusion. In contrast to the two-dimensional case, the upper bound in Eq.~(\ref{Eq27}) remains finite as $B\to0$. Therefore, $I_{3D}$ is finite in this limit, and the long-wavelength Mermin-Wagner argument does not exclude exchange-driven long-range altermagnetic order in three dimensions.

\textit{Discussion---}Although we have focused on a representative $d$-wave altermagnet in the Lieb lattice, the argument is not tied to the microscopic details of the model. Higher-partial-wave altermagnets or other lattice realizations may involve more exchange paths, but short-ranged spin-rotation-invariant exchange interactions still generate only analytic long-wavelength contributions to the Bogoliubov denominator. In two dimensions, such terms cannot remove the logarithmic small-momentum divergence unless an additional spin-symmetry-breaking interaction is present.

In conclusion, we have studied the finite-temperature stability of nonrelativistic altermagnets. Using a two-dimensional $d$-wave altermagnet as a representative example, we found that long-range altermagnetic order is destroyed when the Hamiltonian contains only short-ranged exchange interactions preserving continuous spin-rotation symmetry. In three dimensions, the corresponding small-momentum integral is finite, and long-range altermagnetic order is not forbidden by this argument.
\begin{acknowledgments}
This work was funded by the National Key R$\&$D Program of China (No. 2025YFA1411302 and No. 2022YFA1402802), the National Natural Science Foundation of China (NSFC) (No. 12374103 and No. 12434003), and Sichuan Science and Technology program (No. 2025NSFJQ0045). Z.J. acknowledges financial support from the Alexander von Humboldt postdoctoral fellowship and NSFC (No. 12404125).
\end{acknowledgments}


\begin{thebibliography}{99}
\bibitem{Smejkal3}L. $\rm{\check{S}}$mejkal, J. Sinova, and T. Jungwirth, Beyond Conventional Ferromagnetism and Antiferromagnetism: A Phase with Nonrelativistic Spin and Crystal Rotation Symmetry, \href{https://doi.org/10.1103/PhysRevX.12.031042}{Phys. Rev. X \textbf{12}, 031042 (2022).}
    
\bibitem{Jungwirth2025}T. Jungwirth, R. M. Fernandes, E. Fradkin, A. H. MacDonald, J. Sinova, L. \v{S}mejkal
    , Altermagnetism: An unconventional spin-ordered phase of matter, \href{
https://www.cell.com/newton/fulltext/S2950-6360(25)00154-9}{Newton \textbf{1}, 100162 (2025).}

\bibitem{Smejkal1}L. $\rm{\check{S}}$mejkal, J. Sinova, and T. Jungwirth, Emerging Research Landscape of Altermagnetism, \href{https://doi.org/10.1103/PhysRevX.12.040501}{Phys. Rev. X \textbf{12}, 040501 (2022).}

\bibitem{Hayami2020}S. Hayami, Y. Yanagi, and H. Kusunose, Bottom-up design of spin-split and reshaped electronic band structures in antiferromagnets without spin-orbit coupling: Procedure on the basis of augmented multipoles, \href{https://doi.org/10.1103/PhysRevB.102.144441}{Phys. Rev. B \textbf{102}, 144441 (2020).}

\bibitem{McClarty2024}P. A. McClarty and J. G. Rau, Landau Theory of Altermagnetism, \href{https://doi.org/10.1103/PhysRevLett.132.176702}{Phys. Rev. Lett. \textbf{132}, 176702 (2024).}

\bibitem{Gomonay2024}O. Gomonay, V. P. Kravchuk, R. Jaeschke-Ubiergo, K. V. Yershov, T. Jungwirth, L. $\rm{\check{S}}$mejkal, J. van den Brink, and J. Sinova, Structure, control, and dynamics of altermagnetic textures, \href{https://doi.org/10.1038/s44306-024-00042-3} {npj Spintronics \textbf{2}, 35 (2024).}

\bibitem{Ma2021}H. Ma, M. Hu, N. Li, J. Liu, W. Yao, J. Jia, and J. Liu, Multifunctional antiferromagnetic materials with giant piezomagnetism and noncollinear spin current, \href{https://doi.org/10.1038/s41467-021-23127-7}{Nat. Commun. \textbf{12}, 2846 (2021).}
 
\bibitem{Duan2025}X. Duan, J. Zhang, Z. Zhu, Y. Liu, Z. Zhang, Igor \u{Z}uti\'{c}, and T. Zhou, Antiferroelectric Altermagnets: Antiferroelectricity Alters Magnets, \href{https://doi.org/10.1103/PhysRevLett.134.106801}{Phys. Rev. Lett. \textbf{134}, 106801 (2025).}

\bibitem{Osumi2024}T. Osumi, S. Souma, T. Aoyama, K. Yamauchi, A. Honma, K. Nakayama, T. Takahashi, K. Ohgushi, and T. Sato, Observation of a giant band splitting in altermagnetic MnTe, \href{https://doi.org/10.1103/PhysRevB.109.115102}{Phys. Rev. B \textbf{109}, 115102 (2024).}

\bibitem{Jiang2025}B. Jiang, M. Hu, J. Bai, Z. Song, C. Mu, G. Qu, W. Li, W. Zhu, H. Pi, Z. Wei, Y. Sun, Y. Huang, X. Zheng, Y. Peng, L. He, S. Li, J. Luo, Z. Li, G. Chen, H. Li, H. Weng, and T. Qian, A metallic room-temperature $d-$wave altermagnet, \href{https://www.nature.com/articles/s41567-025-02822-y} {Nat. Phys. \textbf{21}, 754 (2025).}

\bibitem{Smejkal2}L. $\rm{\check{S}}$mejkal, A. Marmodoro, K. Ahn, R. Gonzalez-Hernandez, I. Turek, S. Mankovsky, H. Ebert, S. W. D'Souza, O. $\rm{\check{S}}$ipr, J. Sinova, and T. Jungwirth, Chiral Magnons in Altermagnetic $\rm{RuO_2}$, \href{https://doi.org/10.1103/PhysRevLett.131.256703}{Phys. Rev. Lett. \textbf{131}, 256703 (2023).}

\bibitem{Cui2023} Q. Cui, B. Zeng, P. Cui, T. Yu, and H. Yang, Efficient spin Seebeck and spin Nernst effects of magnons in altermagnets, \href{https://doi.org/10.1103/PhysRevB.108.L180401}{Phys. Rev. B \textbf{108}, L180401 (2023).}

\bibitem{Jin2026} Z. Jin, T. Gong, J. Liu, H. Yang, Z. Zeng, Y. Cao, and P. Yan, Strong Coupling of Chiral Magnons in Altermagnets, \href{https://doi.org/10.1103/g5xq-z15c}{Phys. Rev. Lett. \textbf{135}, 126702 (2026).}

\bibitem{Yang2026} Y. Yang, D. Wang, B. Yang, P. Wang, Y. Mu, Y. Tian, B. Zheng, W. Qin, K. Wang, B. Huang, B. Wang, X. Wan, and D. Wu, Altermagnet-Driven Magnon Spin Splitting Nernst Effect, \href{https://doi.org/10.1103/g5xq-z15c}{Phys. Rev. Lett. \textbf{136}, 026701 (2026).}

\bibitem{Neumann2026}R. Neumann, R. Jaeschke-Ubiergo, R. Zarzuela, L. \v{S}mejkal, J. Sinova, and A. Mook, Odd-Parity-Wave Magnons and Nonrelativistic Thermal Edelstein Effect, \href{https://doi.org/10.48550/arXiv.2603.05415}{arXiv:2603.05415 (2026).}
    
\bibitem{Litvin1974}D. B. Litvin and W. Opechowski, Spin Groups, \href{https://doi.org/10.1016/0031-8914(74)90157-8}{Physica \textbf{76}, 538 (1974).}

\bibitem{Litvin1977}D. B. Litvin, Spin Point Groups, \href{https://journals.iucr.org/paper?S0567739477000709}{Acta Crystallogr. Sect. A \textbf{33}, 279 (1977).}

\bibitem{Mermin1966}N. D. Mermin and H. Wagner, Absence of Ferromagnetism or Antiferromagnetism in One- or Two-Dimensional Isotropic Heisenberg Models, \href{https://doi.org/10.1103/PhysRevLett.17.1133}{Phys. Rev. Lett. \textbf{17}, 1133 (1966).}
    
\bibitem{Wagner1966}H. Wagner, Long-wavelength excitations and the Goldstone theorem in many-particle systems with ``broken symmetries", \href{https://link.springer.com/article/10.1007/BF01325630#citeas}{Z. Physik \textbf{195}, 273 (1966).}
    
\bibitem{Durrnagel2025}M. D\"{u}rrnagel, H. Hohmann, A. Maity, J. Seufert, M. Klett, L. Klebl, and R. Thomale, Altermagnetic Phase Transition in a Lieb Metal, \href{https://doi.org/10.1103/2g3v-z76q}{Phys. Rev. Lett. \textbf{135}, 036502 (2025).}

\bibitem{Kaushal2025}N. Kaushal and M. Franz, Altermagnetism in Modified Lieb Lattice Hubbard Model, \href{https://doi.org/10.1103/s31h-hk2v}{Phys. Rev. Lett. \textbf{135}, 156502 (2025).}
\end{thebibliography}
\end{document}